%% file: main.tex
\documentclass[fleqn,usenatbib,useAMS]{mnras}

\usepackage{tablefootnote}
\usepackage{xcolor}
\usepackage{comment}
\usepackage{amsmath}
\definecolor{color}{RGB}{25,25,112}
\definecolor{negro}{RGB}{0,0,0}
\definecolor{colorurl}{RGB}{25,25,112}
\usepackage{orcidlink}
\usepackage{amssymb}

\usepackage[T1]{fontenc}

\DeclareRobustCommand{\VAN}[3]{#2}
\let\VANthebibliography\thebibliography
\def\thebibliography{\DeclareRobustCommand{\VAN}[3]{##3}\VANthebibliography}

\usepackage{graphicx}	% Including figure files
\title[Photometric Analysis of the Extremely Bright GRB~210822A]{Machine-Learning Enhanced Photometric Analysis of the Extremely Bright GRB~210822A}

\author[Angulo-Valdez et al.]{Camila Angulo-Valdez\,\orcidlink{0009-0002-6667-3294},$^{1}$\thanks{E-mail: camiangulo@astro.unam.mx (CAV)},
Rosa~L.~Becerra\,\orcidlink{0000-0002-0216-3415},$^{2}$
Margarita~Pereyra\,\orcidlink{0000-0001-6148-6532},$^{3}$
Keneth~Garcia-Cifuentes\,\orcidlink{0009-0001-2607-6359},$^{2}$\newauthor
Felipe~Vargas\,\orcidlink{0000-0001-5518-9689},$^{2}$
Alan~M.~Watson\,\orcidlink{0000-0002-2008-6927}$^{1}$
Fabio~De~Colle\,\orcidlink{0000-0002-3137-4633},$^{2}$
Nissim~Fraija\,\orcidlink{0000-0002-0173-6453},$^{1}$\newauthor
Nathaniel~R.~Butler\,\orcidlink{0000-0002-9110-6673},$^{4}$
Maria G. Dainotti\,\orcidlink{0000-0003-4442-8546},$^{5,6,7}$,
Simone~Dichiara\,\orcidlink{0000-0001-6849-1270},$^{8}$
William~H.~Lee\,\orcidlink{0000-0002-2467-5673},$^{1}$\newauthor
Eleonora~Troja\,\orcidlink{0000-0002-1869-7817},$^{9}$
Joshua~S.~Bloom\,\orcidlink{0000-0002-7777-216X},$^{10}$
J.~Jesús~Gonz\'alez\,\orcidlink{0000-0002-3724-1583},$^{1}$
Alexander~S.~Kutyrev\,\orcidlink{0000-0002-2715-8460},$^{11,12}$\newauthor
J.~Xavier~Prochaska\,\orcidlink{0000-0002-7738-6875},$^{13}$
Enrico~Ramirez-Ruiz\,\orcidlink{0000-0003-2558-3102},$^{13}$
and
Michael~G.~Richer\,\orcidlink{0000-0003-4757-1153},$^{14}$
\\
$^1$ Instituto de Astronom{\'\i}a, Universidad Nacional Aut\'onoma de M\'exico, Apartado Postal 70-264, 04510 M\'exico, CDMX, Mexico\\
$^2$ Instituto de Ciencias Nucleares, Universidad Nacional Aut\'onoma de M\'exico, Apartado Postal 70-264, 04510 M\'exico, CDMX, Mexico\\
$^{3}$ CONACYT, Instituto de Astronom{\'\i}a, Universidad Nacional Aut\'onoma de M\'exico, 22860 Ensenada, BC, Mexico\\
$^{4}$ School of Earth and Space Exploration, Arizona State University, Tempe, AZ 85287, USA\\
$^{5}$ National Astronomical Observatory of Japan, 2-21-1 Osawa, Mitaka, Tokyo 181-8588, Japan\\
$^{6}$ Space Science Institute, Boulder, CO, USA\\
$^{7}$ The Graduate University for Advanced Studies, SOKENDAI, Shonankokusaimura, Hayama, Miura District, Kanagawa 240-0193, Japan\\
$^{8}$ Department of Astronomy and Astrophysics, The Pennsylvania State University, 525 Davey Lab, University Park, PA 16802, USA \\
$^{9}$ Department of Physics, University of Rome - Tor Vergata, via della Ricerca Scientifica 1, 00100 Rome, IT\\
$^{10}$ Department of Astronomy, University of California, Berkeley, CA 94720-3411, USA\\
$^{11}$ Department of Astronomy, University of Maryland, College Park, MD 20742-4111, USA\\
$^{12}$ Astrophysics Science Division, NASA Goddard Space Flight Center, 8800 Greenbelt Road, Greenbelt, MD 20771, USA\\
$^{13}$ Department of Astronomy and Astrophysics, UCO/Lick Observatory, University of California, 1156 High Street, Santa Cruz, CA 95064, USA\\
$^{14}$ Instituto de Astronomía, Universidad Nacional Autónoma de México, Unidad Acad\'emica en Ensenada, 22860 Ensenada, BC, Mexico\\
}

\date{Accepted XXX. Received YYY; in original form ZZZ}

\pubyear{2023}

\begin{document}
\label{firstpage}
\pagerange{\pageref{firstpage}--\pageref{lastpage}}
\maketitle

\begin{abstract}
We present analytical and numerical models of the bright long GRB~210822A at $z=1.736$. The intrinsic extreme brightness exhibited in the optical, which is very similar to other bright GRBs (e.g., GRBs 080319B, 130427A, 160625A 190114C, and 221009A), makes GRB~210822A an ideal case for studying the evolution of this particular kind of GRB.
We use optical data from the RATIR instrument starting at $T+315.9$~s, with publicly available optical data from other ground-based observatories, as well as {\itshape Swift}/UVOT, and X-ray data from the {\itshape Swift}/XRT instrument.
The temporal profiles and spectral properties during the late stages align consistently with the conventional forward shock model, complemented by a reverse shock element that dominates optical emissions during the initial phases ($T<300$~s). Furthermore, we observe a break at $T=80000$~s that we interpreted as evidence of a jet break, which constrains the opening angle to be about $\theta_\mathrm{j}=(3-5)$~degrees.
Finally, we apply a machine-learning technique to model the multi-wavelength light curve of GRB~210822A using the {\sc afterglowpy library}. We estimate the angle of sight $\theta_{obs}=(6.4 \pm 0.1) \times 10^{-1}$~degrees, the energy $E_0= (7.9 \pm 1.6)\times 10^{53}$~ergs, the electron index $p=2.54 \pm 0.10$, the thermal energy fraction in electrons $\epsilon_\mathrm{e}=(4.63 \pm 0.91) \times 10^{-5}$ and in the magnetic field $\epsilon_\mathrm{B}= (8.66 \pm 1.01) \times 10^{-6}$, the efficiency $\chi = 0.89 \pm 0.01$, and the density of the surrounding medium $n_\mathrm{0} = 0.85 \pm 0.01~cm^{-3}$.

\end{abstract}


%%%%%%%%%%%%%%%%%%%%%%%%%%%%%%%%%%%%%%%%%%%%%%%%%%

%%%%%%%%%%%%%%%%% BODY OF PAPER %%%%%%%%%%%%%%%%%%

\section{Introduction}

Gamma-ray bursts (GRBs) are the brightest events in the universe \citep{Kumar2015}. It is possible to associate the duration of the event with the progenitors using the parameter $T_{90}$.
Currently, two types of GRBs are known \citep{Kouveliotou1993}. Short GRBs (SGRBs), which are thought to be produced by the merger of two compact objects, typically have $T_{90}\lesssim 2$~s and a harder spectrum \citep{Lee2007,Berger2014}, whereas long GRBs (LGRBs) typically have a $T_{90}\gtrsim 2$~s and a softer spectrum, and are associated with the death of a massive star \citep{Kouveliotou1993,Woosley1993,MacFadyen1999,Hjorth2003,Oates2023}. Nevertheless, in recent years, events with the characteristics of both populations have been reported (see e.g. GRB~200826A \citep{Ahumada2021,Zhang2021}, GRB~211211A \citep{Rastinejad2022,Troja2022}, GRB~210704A \citep{Becerra2023} and GRB~230307A \citep{Levan2023,Yang2023}).

The fireball model \citep{Meszaros1997,Granot2002,Kumar2015} is the most successful theory for interpreting the electromagnetic radiation of GRBs. The model explains the observed prompt (gamma-ray) emission by internal shocks in the jet driven by a central engine \citep{Sari1997} and the late emission or afterglow by the existence of an external shock between the outflow and the circumburst medium \cite[e.g.,][]{Sari1998,Granot2002}. The radiation process that dominates the afterglow is synchrotron emission, produced by relativistic electrons that are continuously accelerated by an ongoing shock interacting with the surrounding medium. This blast wave produces a forward shock that propagates into the surrounding medium and a reverse shock that propagates inward through the relativistic shell \citep{Sari1998,Sari1999}.

For later times, a similar evolution of the X-ray and optical emission would be expected for a GRB \citep{Kumar2015,Zhang2006}. Nevertheless, there are numerous features that appear in optical that are not seen simultaneously in X-rays, such as the reverse shock component and early rebrightening \citep[see e.g.][]{Li2012,Becerra2023b}. These features can suggest different characteristics, such as: the GRB origin being different, they are in a different spectral segment \citep{Granot2002}, are produced by different components, or are produced in a more complex environment \citep[see e.g.][]{Gao2015}. 

Bright GRBs permit follow-up for a longer period of time \citep[see e.g.][]{OConnor2023,Becerra2023,Becerra2019b}, leading to a better understanding of their progenitors and circumburst environments. Given that only about 40\% of these phenomena have a well-determined distance \citep{Evans2007,Evans2009}, the study of particularly bright GRBs with known distances is extremely valuable because they allow us to explore in more detail the morphology and structure of the jet \citep[see e.g.][]{Gill2019}. 

%These ideas are the main motivation for this paper. 
Here, we present photometric observations of the long GRB~210822A in the optical, UV, and X-rays during the afterglow phase and compare them with the detailed predictions of the fireball model.

Our paper is organised as follows. In Section~\ref{sec:observations}, we present observations with {\itshape Swift}, RATIR, Katzman Automatic Imaging Telescope (KAIT), the Nordic Optical Telescope (NOT), and other ground-based telescopes. In Section~\ref{sec:analysis} we present our analysis. We discuss the nature of the GRB~210822A and its interpretation within the context of the current \emph{fireball} model, and we conduct a comparative analysis of our photometry with other events and deduce the physical parameters that could describe this explosion in Section~\ref{sec:discussion}. Finally, we provide a summary of our results in Section~\ref{sec:summary}.

\section{Observations}
\label{sec:observations}

\subsection{High-energy}
\label{sec:swift}

The {\itshape Swift}/BAT instrument triggered on GRB~210822A at $T = $ 2021 August 22 09:18:18 UTC \citep{30677}. 
%from the bat refiend analysis:
The {\itshape Swift}/BAT on-board location was 20:17:50.7 +05:15:54.9 J2000, with an uncertainty of 3.0~arcmin (radius, 90\% containment, including systematic uncertainty). The {\itshape Swift}/BAT mask-weighted light curve showed a multi-peaked structure that started and peaked at the time of trigger ($T$). The main pulse structure ended at about $T+ 20$~s, followed by a long tail that lasted until about $T+400$~ s. The duration of GRB~210822A  (15-350~keV) was $T_{90}=185.8\pm 46.6$~seconds, making it a long GRB, with an estimated fluence of 2.0$\pm$0.04$\times10^{-6}$~erg~cm$^{-2}$ in the 15--150~keV band \citep{30689}.
We investigate the similarities of GRB~210822A with previous {\itshape Swift}/BAT events \citep{Evans2007,Evans2009}, in order to identify other similar phenomena using a machine learning approach. The method and results are presented in Section~\ref{sec:gammasim}. 

The Konus-Wind instrument reported observations %from which they produced a light curve with a multi-peaked pulse 
starting at about $T-1$~s and with a total duration of about 12~s. According to \cite{30694}, the burst had a fluence of $1.20 \pm 0.11 \times~10^{-4}$ erg~cm$^{-2}$ and a 64 ms peak flux, measured from $T+0.320$~s, of $3.25 \pm 0.28 \times~10^{-5}$ erg~cm$^{-1}$ (both in the 20~keV to 10~MeV energy range). 

The {\itshape Swift}/X-ray Telescope (XRT) instrument started observing the field at $T+74.6$~s and found an uncatalogued X-ray source at 20:17:45.01 +05:17:00.6 J2000 with an uncertainty of 2.0~arcsec (radius, 90\% containment). Its location, only 145~arcsec from the {\itshape Swift}/BAT on board position \citep{30677}, was well inside the {\itshape Swift}/BAT error circle.  

Finally, the {\itshape Swift}/Ultra-violet Optical Telescope (UVOT) instrument began settled observations of the field of GRB~210822A at $T+85$~s, detecting a fading source consistent with the {\itshape Swift}/XRT position \citep{30710}. {\itshape Swift}/UVOT took a finding chart exposure of 250~seconds with the $u$ filter starting at $T+295$~s \citep{30677}. The afterglow candidate was found in the initial data products in all filters.

\subsection{Optical}

\subsubsection{RATIR}
\label{sec:observationsratir}
The Reionization and Transients InfraRed camera (RATIR)\footnote{\url{http://ratir.astroscu.unam.mx/}} is a four-channel simultaneous optical and near-infrared imager. It was mounted on the 1.5~meters Harold L. Johnson Telescope at the Observatorio Astron\'omico Nacional in Sierra San Pedro M\'artir in Baja California, Mexico, until summer 2022. RATIR autonomously responded to GRB triggers from the Swift satellite and obtained simultaneous photometry in $riZJ$ or $riYH$ bands \citep{Butler2012,Watson2012,Littlejohns2015,Becerra2017}. The reduction pipeline performed bias subtraction and flat field correction, followed by astrometric calibration using \href{astrometry.net}{\sc astrometry} software \citep{Lang2010}, iterative sky subtraction, co-addition using {\sc SWARP}, and source detection using {\sc sextractor} \citep{Bertin1996,Becerra2017,Pereyra2022}. We calibrated against the U.S. Naval Observatory B1 catalogue (USNO-B1) \citep{Monet2003} and the Two Micron All-Sky Survey (2MASS) \citep{Skrutskie2006}.

RATIR observed the field of GRB~210822A from 2021 August 22 09:19:30 to 09:48:54 UTC (from $T + 0.02$ to $T + 0.49$~hours after the BAT trigger), obtaining a total of 0.19~hours of exposure in the $r$ and $i$ bands \citep{30680}. RATIR obtained individual frames of 80~s of exposure time. 
RATIR photometry is presented in Table~\ref{tab:observations}, as magnitudes in the AB system and without correction for the Milky Way extinction in the direction of the burst, and shown in Figure~\ref{fig:reverse}.

\input{observations_RATIR}

\subsubsection{Other optical ground-based facilities}

\begin{figure*}
\includegraphics[clip, width=0.9\textwidth]{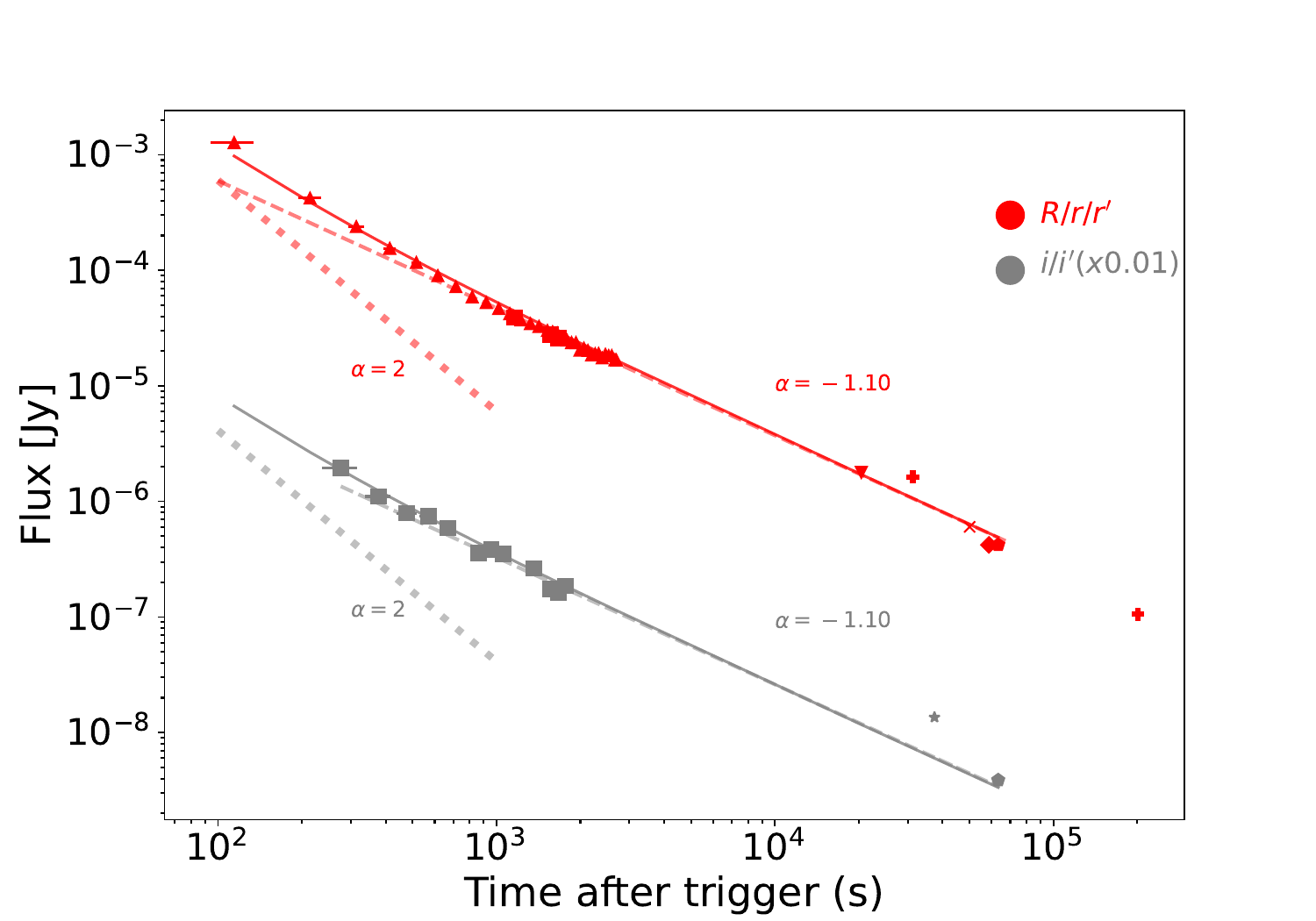}
 \caption{Optical light curves of GRB~210822A. The red and gray symbols correspond to data in the \emph{r} and \emph{i} bands, respectively. Different symbols indicate data from RATIR (squares), KAIT (upward triangles), GROND (pentagons), NOT (crosses), TSHAO (pluses), CAHA (stars), and Abbey Ridge (diamonds). We illustrate the reverse shock (dotted line) and forward shock (dashed line) components that best fit the optical observations, as well as the sum of both (continuous line), for both the $r$ and $i$ bands. The temporal slopes that better describe the forward shock emission during the afterglow are also shown. The evolution of the light curve after $T+80000$~s is discussed in Section~\ref{sec:jetbreak}}
 \label{fig:reverse}
\end{figure*}

We complement RATIR photometry with data from other optical telescopes from the GCN/TAN circulars, such as the 0.76~m Katzman Automatic Imaging Telescope (KAIT) \citep{30679}, the NEXT-0.6~m telescope (NANSHAN)  \citep{30684}, the Nordic Optical Telescope (NOT) \citep{30692}, the Gamma-Ray Burst Optical/Near-Infrared Detector (GROND) \citep{30703}, the Tian Shan Astronomical Observatory (TSHAO)  \citep{30712}, the Calar Alto Astronomical Observatory (CAHA) \citep{30723} and the 0.35~m telescope at the Abbey Ridge Observatory \citep{30701}. This photometry is listed in Table~\ref{tab:observations} and shown in Figure~\ref{fig:reverse}.

 \cite{30692} estimated a redshift of $z=1.736$ for the GRB~210822A using spectroscopic observations with NOT in the $3700-9500$~{\AA} range that showed a continuum with absorption characteristics from the \ion{Fe}{II}, \ion{Mg}{II}, \ion{Mg}{I} and \ion{Al}{III} lines.

Additionally, \citet{30677} and \citet{30710} reported a reddening of $E(B-V) = 0.162$ in the direction of the burst. We calculate the extinction $A_r$, adopting the relationship $A_V=3.1\times$$E(B-V)$ \citep{Schlegel1998} and $A_r/A_V\approx0.9$ \citep{Gordon2003}. 

\section{Temporal and spectral analysis}
\label{sec:analysis}

In the framework of the fireball model, the afterglow phase is explained as synchrotron radiation from  a population of accelerated electrons $N$ with a power-law energy distribution $N(E)\propto E^{-p}$, where $E$ and $p$ are the energy and index respectively. Typically, the observed flux density $F_\nu$ has a multisegmented power-law profile in both time $t$ and frequency $\nu$, denoted as $F_{\nu}\propto t^{-\alpha}\nu^{-\beta}$ \citep{Sari1998,Granot2002}.

\subsection{Temporal Analysis}
\label{sec:temporalanalysis}

Figure~\ref{fig:reverse} shows the best fit to our optical photometry. Our data, in both filters, are well-modelled by one single power law segment for times between $T+300$~s a  $T+80000$~ with a power law $F_\nu\propto t^{-1.10\pm 0.01}$. This finding is in good agreement with the average decay index observed in large samples of GRB afterglows \citep{Liang2010,Oates2009} and is consistent with an isotropic forward shock in a constant-density medium. However, we note that for early times our photometry in the $r$ band is better fit when a reverse-shock component is considered. (The data in the $i$ band do not extend to such early times, but are also consistent with a reverse shock.) 
For a detailed discussion on the temporal evolution of the afterglow we refer to Sections~\ref{sec:early} and \ref{sec:jetbreak}. 

\subsubsection{Early Afterglow}
\label{sec:early}

Figure~\ref{fig:reverse} shows the best fit to our photometry. Both optical filters are well-modelled by one single power law segment for times after $T+300$~s. We observe that for early times ($T<300$~s), the photometry in the $r$ band lies above our fit. To address this discrepancy, we fit these points independently using a distinct power law (depicted by the red dotted line in Figure~\ref{fig:reverse}). We consider a function $F_\nu\propto t^{-2}$ according to the anticipated decay pattern of a reverse shock component \citep{Gao2015}, obtaining a better match to our photometry. We interpret this feature as the signature of the reverse shock propagating in the inner shell of the relativistic jet present in GRB~210822A \citep{Sari1999,Becerra2019b,Becerra2019c,Gomboc2009,Meszaros1999}. The presence of a reverse shock suggests the presence of a matter-dominated shell, which implies a low level of magnetisation \citep{Zhang2005,Zhang2011,GarciaGarcia2023}.

%Although the lack of early times photometry in another optical band makes impossible a more rigorous comparison at these wavelenghts, based on the {\itshape Swift}/XRT light curve, we can confirm that no contribution of this emission component is observed in X-rays. % instrument makes it possible to ensure that we do not observe a counterpart for this excess in X-rays. 

\subsection{Spectral Analysis}
\label{sec:spectrum}

In order to complement our photometric analysis, we retrieved light curves and spectra from the UK Swift Science Data Centre (UKSSDC) online repository \citep{Evans2009} in the range of 0.3--10~keV. We binned all data retrieved from Swift/XRT, for Window Timing and Photon Count modes independently \citep[see][]{Evans2007}, considering groups of 3 consecutive exposures, which yields a variable size for the time bins. From $T + 78$~s to $T + 1300$~s, early times, bin sizes are around 1 to 7~s. The size bin increases significantly to tens or even hundreds of seconds for the late time observations, from $T + 4500$~s to $T + 502000$~s, where data are sparse. 

\begin{figure}
\centering
\begin{tabular}{cc}
\includegraphics[width=1\linewidth]{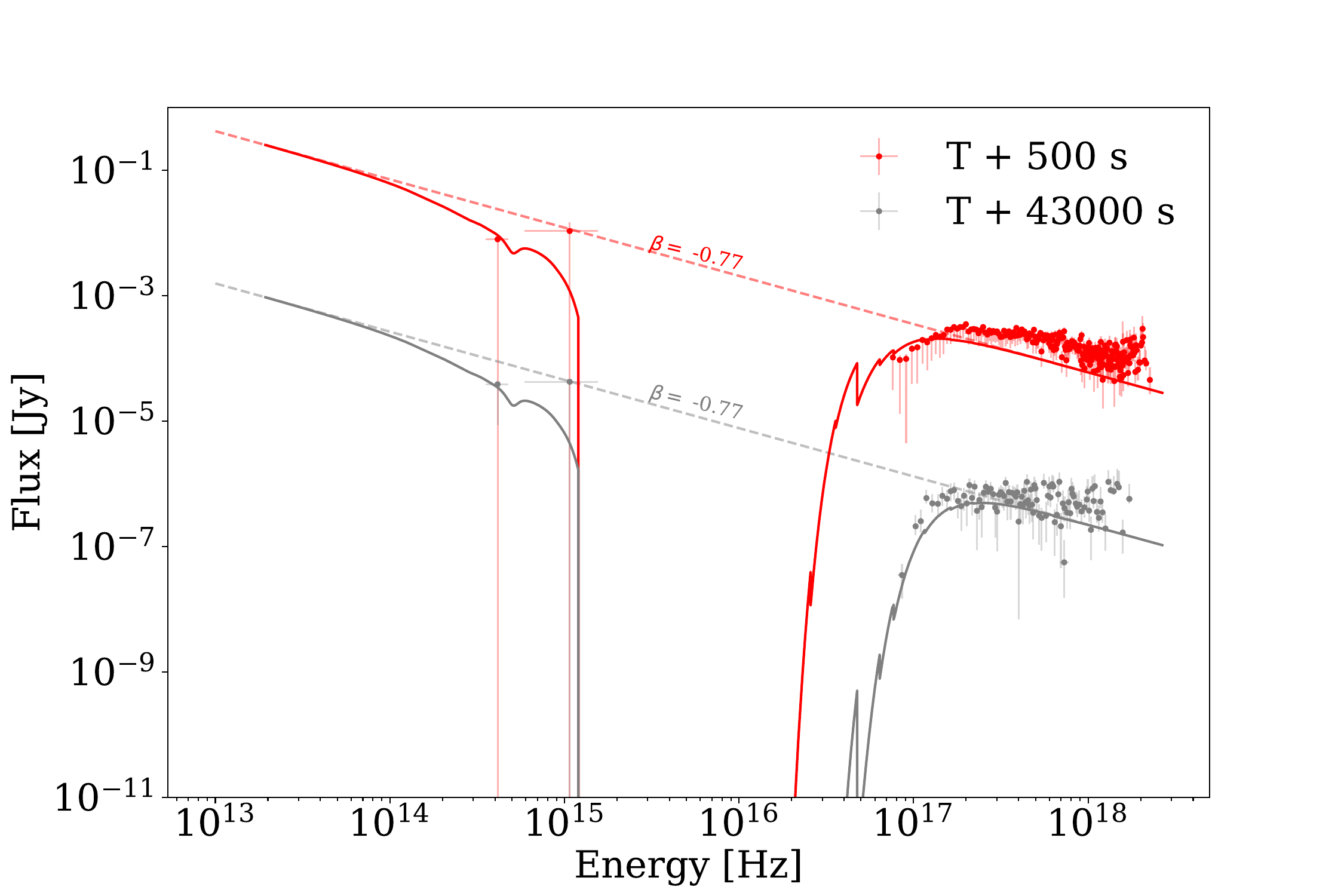}
\end{tabular}
 \caption{SED of GRB~210822A at $T + 500~s$ (red) and $T + 43 000~s$ (gray). We use optical data from Table~\ref{tab:observations} and {\itshape Swift}/XRT data retrieved from the UKSSDC. We show the corresponding {\sc Xspec} model (continuous lines), which considers the effects of reddening and absorption by the dust, as well as the linear fit (dashed lines).}
 \label{fig:sed}
\end{figure}

The optical and X-ray data at $T+500$ and $T+43000$~s are well-fitted with a simple absorbed power law of slope $\beta=-0.77\pm 0.03$ (see Figure~\ref{fig:sed}), which indicates that they belong to the same spectral segment. $F_\nu\propto\nu^{-0.77\pm 0.03}$ yields an electron index of $p=2.54 \pm 0.10$ for GRB~210822A, and implies a value of $F_\nu\propto t^{-1.15\pm 0.06}$  that is consistent with the temporal decay found in Section~\ref{sec:temporalanalysis}. Compared to the expected afterglow spectrum detailed in \citet{Granot2002}, we identify this segment as the frequency interval where $\nu_\mathrm{m}<\nu_\mathrm{opt}<\nu_{X}<\nu_c$.

Figure~\ref{fig:sed} also shows the fit of the model {\sc xspec} that considers the effects of reddening and absorption by dust \citep{Arnaud1996}.  
We performed our analysis at $T+500$~s and $T+43000$~s to compare the evolution of the SED. We used a reddening of $E(B-V)=0.162$ \citep{30677}, a redshift of $z=1.786$ \citep{30692}, and a column density of $1.45\times10^{21}$cm$^{-2}$ \citep{30691}. We obtained reduced $\chi^2=3.07$ and $\chi^2=0.92$ with 698 and 300 degrees of freedom, at $T+500$~s and $T+43 000$~s, respectively.

\subsection{Machine Learning Method}
\label{sec:ML}
Machine learning (ML) has found diverse applications within astrophysics, spanning from exoplanet identification \citep{Malik22} and galaxy categorisation \citep{Ferreira20} to supernova classification \citep{Lochner16}, spectral modelling \citep{Vogl20}, and gravitational wave detection \citep{Ni16}. It is worth noting that the majority of these applications consider ML for classification purposes, with limited utilisation in modelling.

\begin{figure}
\includegraphics[clip, width=1\linewidth]{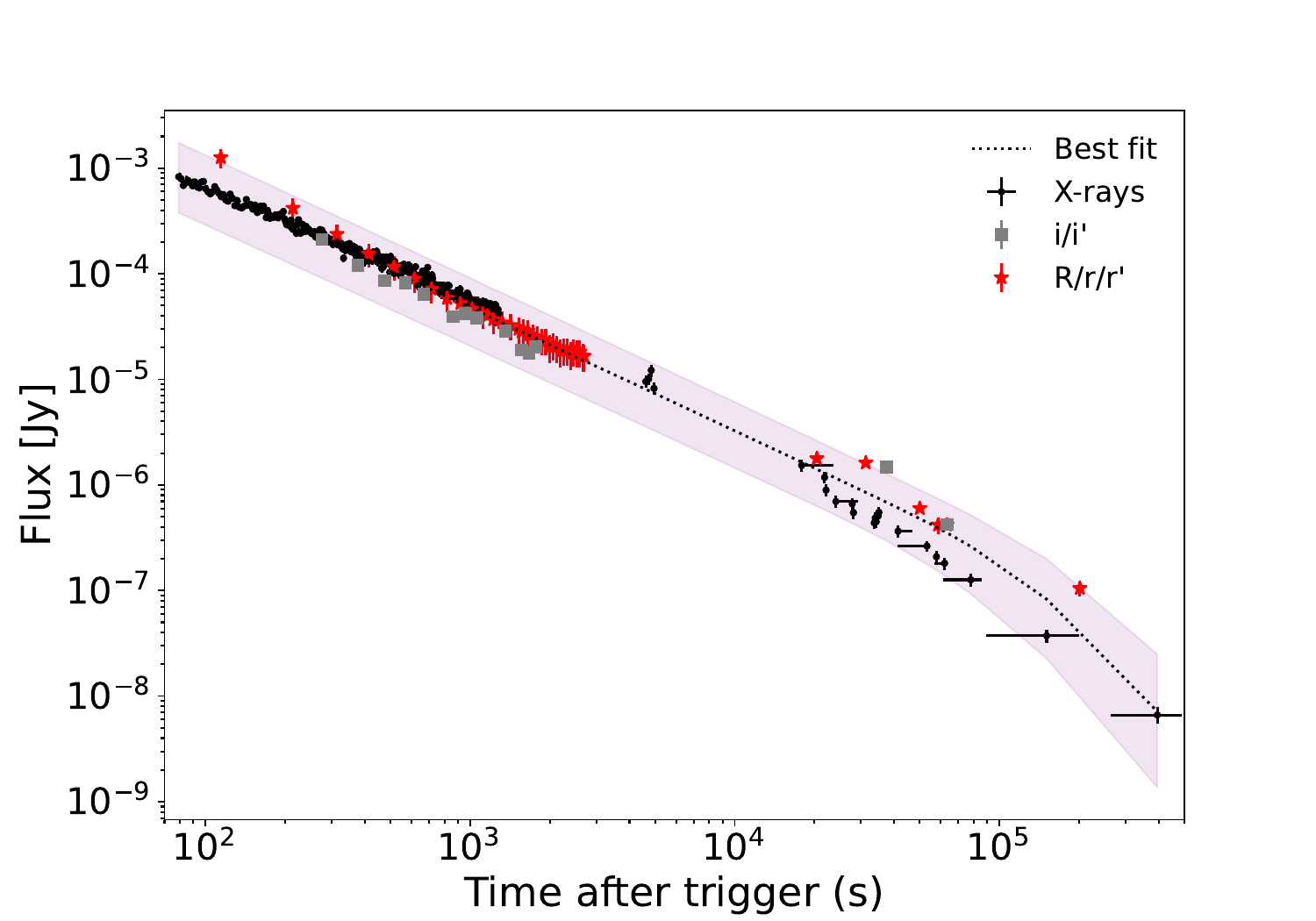}
 \caption{X-ray and optical light curves of GRB~210822A. Optical photometry is re-scaled to a common frequency of 1~keV by employing the scaling $\nu^{-0.77}$ obtained in Section~\ref{sec:spectrum}. We illustrate the model of the light curve of GRB~210822A using ML (dashed black line) and the $1\sigma$ error region (shaded magenta area). The parameters of our best fit are listed in Table~\ref{table:parameter_values}.}
 \label{fig:lightcurve}
\end{figure}

Based on the results obtained in the Section~\ref{sec:spectrum}, we plot in Figure~\ref{fig:lightcurve} the light curves in the $r$ and $i$ bands by scaling the optical flux to a common frequency of 1~keV, considering $\nu^{-0.77}$, and employ ML methodologies to build a predictive model for the afterglow of GRB~210822A.

Details of this implementation are described in Appendix~\ref{sec:neuronal}. To effectively train and evaluate the ML model, we generate synthetic light curve data with the {\sc afterglowpy library} \citep{Ryan2020}. This synthetic data set is then used to train the ML model built in the Keras and TensorFlow libraries, as referenced in \cite{keras15}, which subsequently processes the actual observational photometric data from Table~\ref{tab:observations} to extract the parameters that align with the prediction of the model, introducing a novel and unprecedented approach in the field. 

We created a diverse data set comprising 30,000 synthetic GRB light curve models. These models included values of random parameters within predefined ranges for the angle of view $\theta_{obs}$, the energy $E_0$, the opening angle $\theta_\mathrm{j}$, the electron index $p$, the thermal energy fraction in electrons $\epsilon_\mathrm{e}$ and in the magnetic field $\epsilon_\mathrm{B}$, efficiency $\chi$, and the density of the  surrounding medium $n_\mathrm{0}$ \citep{Sari1998,Granot2002}. These parameters are listed in Appendix~\ref{table:parameter_ranges}. 

The trained ML model was evaluated with observed GRB data, corrected by Galactic extinction, producing the parameter estimates shown in Table \ref{table:parameter_values}. The results of the optimal parameters applied to {\sc afterglowpy} can be seen in Figure~\ref{fig:lightcurve}.

Nevertheless, we note that a machine learning method such as the one employed in the fit presented in Figure ~\ref{fig:lightcurve}, inherently selects a single set of parameters. The 1$\sigma$ error region is used to illustrate how well the best model fits the data, for the selected parameters (see details in Appendix~\ref{sec:neuronal}). As discussed by \cite{GarciaCifuentes2023b}, in several adjustment methods, to find degeneracy in the parameters is not unusual. Such is the case of the relationship (scaling function) found between $E_0$, $n_0$ and $\epsilon_\mathrm{B}$ depending on the value of $p$, in the same spectral regime for the afterglow \citep{vanEerten2012,Granot2012,GarciaCifuentes2023b}. Thus, the photometry of the GRBs can be reproduced by a range of values given their energy, density and microphysical parameters.

\begin{table}
\centering
\begin{tabular}{|c|c|}
\hline
\hline
Parameter & Value $\pm$ Error \\
\hline
\hline
%$\theta_{obs}$ & $(1.12 \pm 0.013) \times 10^{-2}$~rad \\
$\theta_\mathrm{obs}$ & $(6.4 \pm 0.1) \times 10^{-1}$~degrees \\
$E_0$ & $(7.9 \pm 1.6) \times 10^{53}$~ergs \\
%$\theta_{j}$ & $(6.4 \pm 0.93) \times 10^{-2}$~rad\\
$\theta_\mathrm{j}$ & $(3.7 \pm 0.5)$~degrees\\
$p$ & $2.54 \pm 0.10$ \\
$\epsilon_\mathrm{e}$ & $(4.63 \pm 0.91) \times 10^{-5}$ \\
$\epsilon_\mathrm{B}$ & $(8.66 \pm 1.01) \times 10^{-6}$ \\
$\chi$ & $0.89 \pm 0.01$ \\
$n_\mathrm{0}$ & $(0.85 \pm 0.01)$~cm$^{-3}$ \\
\hline
\end{tabular}
\caption{Parameter Values and Errors}
\label{table:parameter_values}
\end{table}

\section{Discussion}
\label{sec:discussion}

%To constrain the parameters which describe the jet of GRB~210822A, we detail the physical implications of the temporal indices reported in Section~\ref{sec:temporalanalysis} framed within the context of the Fireball model.

\subsection{Jet break}
\label{sec:jetbreak}

From Figure~\ref{fig:lightcurve}, we suggest the presence of a jet break between $T+80000$~s and $T+100000$~s (0.93--1.16 days). Considering this scenario, the opening angle for a constant density ISM depends on the jet break time as:

\begin{eqnarray}\nonumber
\theta_\mathrm{j}&=&
0.07 \textrm{ rad}{\left(\frac{t_\mathrm{break}}{1 \textrm{ day}}\right)}^{3/8}
{\left(\frac{1+z}{2}\right)}^{-3/8}\\
%\nonumber
&\times&\textrm{ }{\left(\frac{E_\mathrm{K,iso}}{10^{53}\textrm{ erg}}\right)}^{-1/8}{\left(\frac{n}{0.1\textrm{ 
 cm}^{-3}}\right)}^{1/8}
\end{eqnarray}

where, $E_\mathrm{K,iso}$ is the isotropic kinetic energy of the blast wave and $n$ refers to the density of the circumburst medium \citep{Wang2018}. Assuming typical values of $n_0=0.85$~cm$^{-3}$ (see Section~\ref{sec:ML}) and $E_\mathrm{K,iso}$=10$^{53}$--10$^{55}$~erg, and the measured value of $z=1.74$, we obtain an opening angle of the jet $\theta_\mathrm{j}=3$--$5\pm 0.5$~degrees.
Our estimated opening angle is consistent with the lower limits for the existing GRB sample \citep{Berger2014} and in good agreement with the value obtained from machine-learning modelling in Section~\ref{sec:ML}.

\subsection{GRB~210822A in the context of other GRBs}

\subsubsection{Gamma-rays}
\label{sec:gammasim}

% machine learning methods have been used as an optimal way to classify GRBs by their light curve features.

The classification of extremely bright LGRBs is essential for understanding their nature and origin. Recently, machine-learning methods provide a faster and more consistent way to classify transients and, specifically, GRBs, by automatically learning from the data and identifying patterns that may be difficult for humans to detect. Furthermore, machine-learning methods can handle large amounts of data, including multi-wavelength and multi-messenger observations, which are crucial for a comprehensive understanding of GRBs \citep[see e.g.][]{Jespersen2020, GarciaCifuentes2023,Steinhardt2023}. 

Therefore, following the method presented by \cite{GarciaCifuentes2023} using the {\sc ClassiPyGRB} library\footnote{\url{https://github.com/KenethGarcia/ClassiPyGRB}}. We analyse the {\itshape Swift}/BAT catalogue in order to find those GRBs whose gamma-ray emission measured by the {\itshape Swift}/BAT instrument are similar to GRB~210822A. This method is based on a popular non-linear dimensionality algorithm called t-SNE, which transforms high-dimensional data (in this case the {\itshape Swift}/BAT GRB light curves) into a 2-dimensional representation while preserving pairwise similarities \citep{vandermaaten08}. t-SNE ensures that the distances between points in the lower-dimensional representation reflect the similarities between patterns. For this work, we compare the discrete-time Fourier transform of the photometry in all bands of in the original {\itshape Swift}/BAT data.
As demonstrated in \cite{GarciaCifuentes2023}, we reduce the noise impact by using the non-parametric approach presented by \citet{FABADA22} on 1451 {\itshape Swift} / BAT light curves. Then, we perform the following procedure:

\begin{itemize}
    \item The light curves binned at $64$ ms are filtered using the $T_{100}$ duration presented in the {\itshape Swift}/BAT GRB catalogue.
    \item We normalised the light curves by the total fluence and ensured that every GRB has the same number of data points by padding zeros after the trigger in all bands.
    \item We perform a discrete-time Fourier transform (DTFT) on all events.
    \item We applied the t-SNE method to the Fourier Amplitude Spectrum for the entire sample.
\end{itemize}

\begin{figure*}
\centering
\begin{tabular}{cc}
\includegraphics[width=0.9\textwidth]{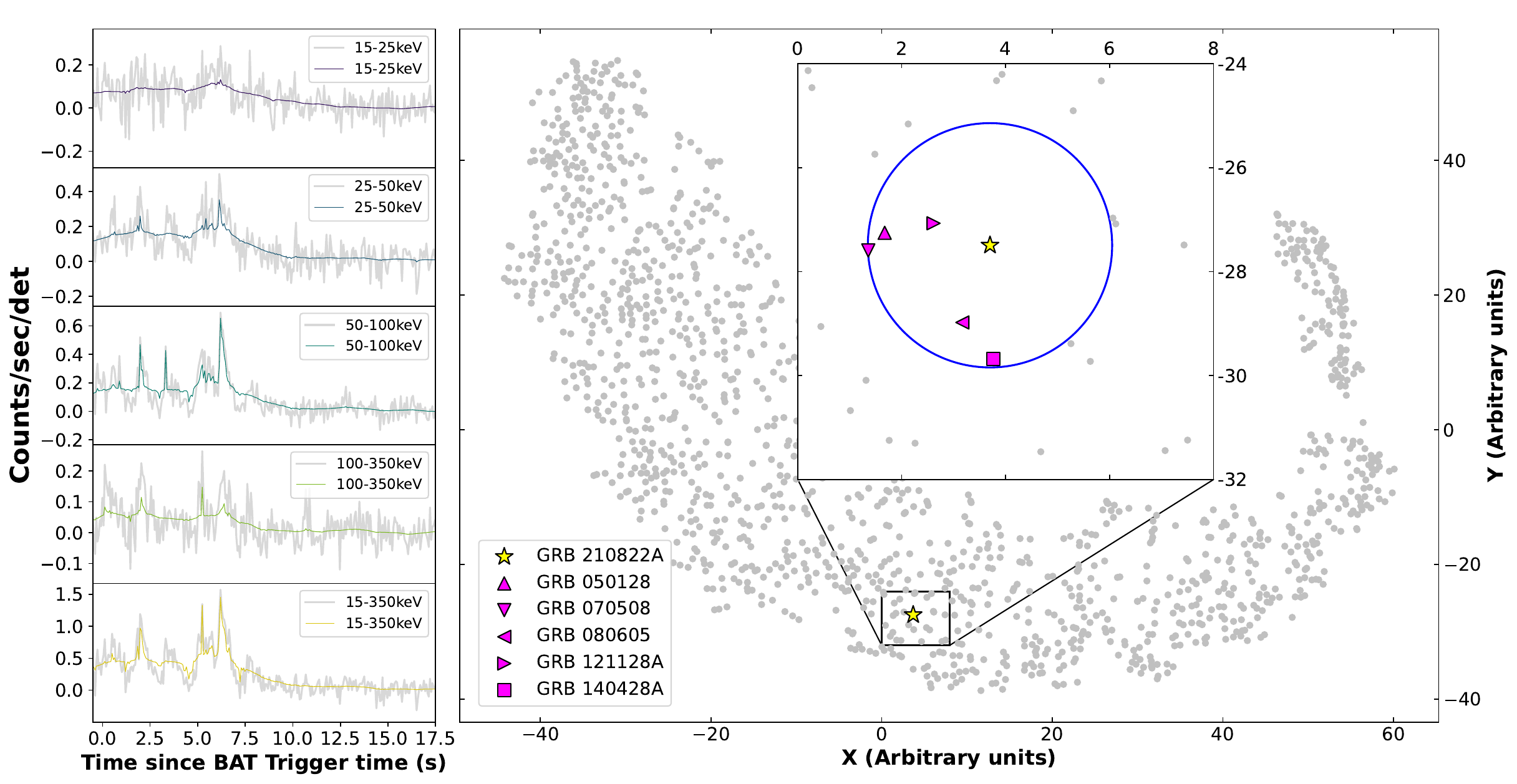}
\end{tabular}
 \caption{t-SNE visualization of the {\itshape Swift}//BAT data light curves using the method described by \citet{GarciaCifuentes2023}. The light curve obtained in the four bands of GRB~210822A by {\itshape Swift}/BAT in its automatic reduction pipeline is illustrated in the left panel. The blue circle indicates the equidistant region where the 5 closest events to GRB 210822A are located.}
 \label{fig:ml}
\end{figure*}

In Figure~\ref{fig:ml} we show all {\itshape Swift}/BAT GRBs on the t-SNE map. In this lower-dimensional plane, points that are closer to each other have more similarities in their original data than points that are further apart. This means that t-SNE maps provides a visual representation of the whole dataset, where the relative distances in the parameter space encode the relationships and similarities between the {\itshape Swift}/BAT light curves of the GRBs, allowing an intuitive understanding of their patterns in the dataset \citep{vandermaaten08}. In the t-SNE plot, we constrain the neighbourhood of GRB~210822A to a circle subtended by the radius that contains the five closest data points.\footnote{We have chosen to analyse the five nearest neighbours. Nevertheless, this analysis is flexible and can be repeated with a variable number of neighbours $n$ based on the specific research requirements.}
. The blue circle in the inset of Figure~\ref{fig:ml} indicates this region, and the nearest neighbors to GRB210822A are also listed (GRB050128, GRB080605, GRB080319C, GRB121128A, and GRB140428). We present some of their physical properties in Table~\ref{tab:similar}.

\begin{table}
	\centering
	\caption{GRBs with similar {\itshape Swift}/BAT light curves to those of GRB~210822A. References: (1) \citet{Xiao2011}, (2) \citet{6390}, (3) \citet{6398}, (4) \citet{Fynbo2009}, (5) \citet{14011}, (6) \citet{14009}, (7) \citet{16186}, (8) \citet{16181}.} 
	\label{tab:similar}
 \begin{tabular}{lrcrc}
		\hline
		\hline
			GRB & $T_{90}$ [s] & Redshift & References & t-SNE \\ 
			& & & &  distance $^{\rm \dagger}$ \\ 
		\hline
		\hline
    \input{similar}
	\end{tabular}
 \footnotesize{$^{\rm \dagger}$ Referenced to the arbitrary units used in Figure~\ref{fig:ml}. }
\end{table}

In the next Section, we investigate the similarities in the optical counterpart and we will compare the events that are similar in that band with the set of similar GRBs in gamma-rays presented in this section and the implications of our results.

\subsubsection{Optical}
\label{sec:opticaluvot}
\begin{figure*}
\begin{tabular}{cc}
    \includegraphics[width=0.9\textwidth]{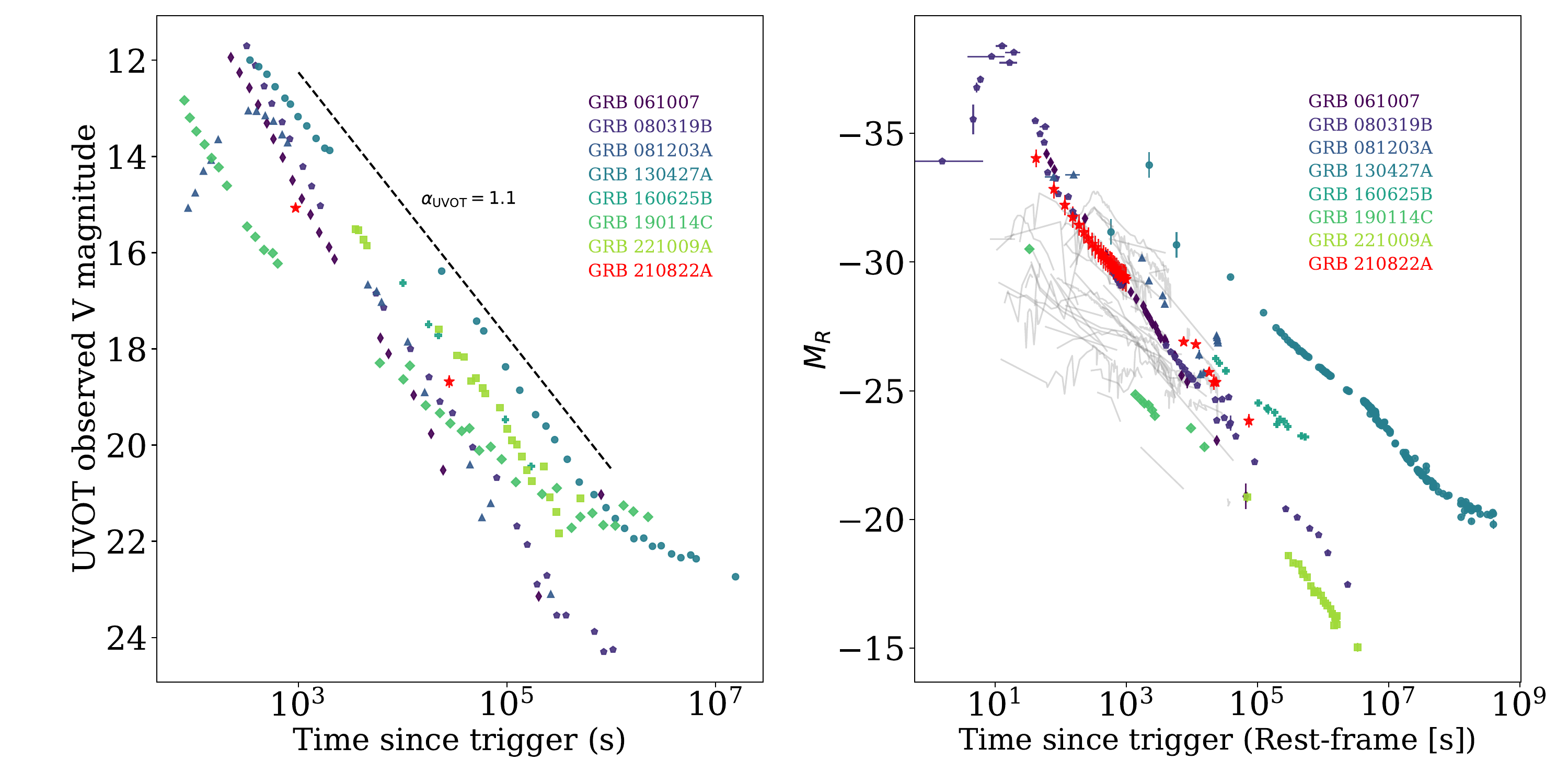}
\end{tabular}
    \caption{Comparison of optical light curves of GRB~210822A (red stars) and the brightest long GRBs presented by \citet{Oates2023}: GRB~061007 (thin diamonds, \citet{Mundell2007}), GRB~080319B (pentagons, \citet{Bloom2009}), GRB~081203A (up triangles, \citet{8645,8604,8617}), GRB~130427A (filled circles, \citet{Perley2014,Fraija2016}), GRB~160625B (crosses, \citet{Troja2017,19602,19599,19605,19619,19640,19642,19651}), GRB~190114C (diamonds, \citet{Fraija2019}), and GRB~221009A (squares, \citet{Fulton2023,OConnor2023}). 
    \emph{Left}: Optical light curves in the observer's frame with the {\itshape Swift}/UVOT instrument in the V band.
    We plot a power-law segment that illustrates the typical decay of $\alpha_\mathrm{UVOT}=1.1$. \emph{Right}: Optical light curves of the same set of GRBs and placed at their corresponding rest frame. We complement the sample with the GRBs with known redshift presented in \citet{Becerra2023b}.}
    \label{fig:uvot_r}
\end{figure*}

In the left panel of Figure~\ref{fig:uvot_r}, we compare the photometry of GRB~210822A in the $v$ filter obtained by the {\itshape Swift}/UVOT instrument \citep{30710} with the current sample of remarkably bright long GRBs with {\itshape Swift}/UVOT light curves, presented by \cite{Oates2023}. This sample includes GRBs with unusual behaviour such as the ultra-bright GRB~061007, GRB~080319B, GRB~081203A, GRB~130427A, GRB~160625B, GRB 190114C, and GRB~221009A, all of which are classified as spectacular GRBs in terms of their brightness, when compared with other 56 UVOT-observed GRBs discovered between 2005 and 2010 \citep{Oates2012}.

We look at the $r$ band photometry for the same set of GRBs, presented in the left panel of Figure~\ref{fig:uvot_r}.  We use the data from \citet{Mundell2007} (GRB~061007 , ), \citet{Bloom2009}(GRB~080319B), \citet{8645, 8604, 8617} (GRB~081203A),  \citet{Perley2014, Fraija2016} (GRB~130427A),  \citet{Troja2017, 19602, 19599, 19605, 19619, 19629, 19680, 19640, 19642, 19651} (GRB~160625B),  \citet{Fraija2019} (GRB~190114C), \citet{Fulton2023, OConnor2023} (GRB~221009A). We also plotted the optical light curves of standard GRBs from \citet{Becerra2023b} (gray solid lines) for comparison. For the estimation of the distances, we assume a $\Lambda$CDM model with a $H_0=67.8$ km/Mpc/s \citep{PlanckCollab2016}.

\begin{table*}
	\centering
	\caption{Physical properties of the brightest {\itshape Swift}/UVOT long GRBs  identified by \citet{Oates2023} and the GRB~210822A. Information about duration and redshift is from the BAT catalogue. References: 
%%% 061007
    1. \citet{5713}, 2. \citet{Fynbo2009}, 3. \citet{Mundell2007}, 4. \citet{Schady2008}, 5. \citet{Panaitescu2020}, 6. \citet{5722}, 
 %%%% 080319B
    7. \citet{7444},  8. \citet{Racusin2008}, 9. \citet{Tanvir2010}, 10. \citet{Pandey2009}, 11. \citet{7482}, 
 %%%% 081203A
    12. \citet{Kuin2009}, 
 %%%% 130427A
    13. \citet{14455}, 14. \citet{Perley2014}, 15. \citet{Maselli2014}, 16. \citet{Laskar2013}, 17. \citet{Liu2013}, 18. \citet{14487}, 
 %%%% 160625B
    19. \citet{19587} 20. \citet{19600}, 21. \citet{Strausbaugh2019}, 22. \citet{Lü2017}, 23. \citet{Alexander2017}, 24. \citet{Fraija2017}, 25. \citet{19604}, 
 %%%% 190114C
    26. \citet{23695}, 27. \citet{Misra2021}, 28. \citet{Campana2021}, 29. \citet{23737}, 
 %%%% 221009A
    30. \citet{32688}, 31. \citet{Laskar2023}, 32. \citet{An2023}, 33. \citet{LHAASO2023}, 34. \citet{Negro2023}, 35. \citet{Frederiks2023}, 
 %%%% 210822A
    36. \citet{30689}, 37. \citet{30692}, 38. \citet{30694}, 39. This work.}
    
	\label{tab:uvotbright}
 \begin{tabular}{lrccccr}
		\hline
		\hline
			Event & $T_{90}$ [s]  & $E_\mathrm{iso}$ [erg] & $L_\mathrm{iso}$[erg/s] & z & $\theta_\mathrm{j}$ [$^\circ$] & References \\ 
		\hline
		\hline
    \input{uvot_r}
    $^{\rm *}${\itshape Fermi}/GBM GRB
	\end{tabular}
 %\footnotesize{$^{\rm *}${\itshape Fermi}/GBM GRB}
\end{table*}

According to Figure~\ref{fig:uvot_r}, it can be seen that GRB~210822A has many similarities with the population of intrinsically bright GRBs (with $E_\mathrm{iso}\sim10^{54}$ erg), sharing not only qualitative similarities in their temporal evolution, such as the observed slope and fluxes, but also their low redshift and opening angle $\theta_\mathrm{j}$ (see Table~\ref{tab:uvotbright}). 
Moreover, GRB~210822A shares the presence of a reverse shock component with GRB~080319B \citep{Bloom2009}, GRB~130427A \citep{Perley2014}, GRB~160625B \citep{Zhang2018,Fraija2017}, GRB~190114C \citep{Laskar2019,Fraija2019} and GRB~221009A \citep{Laskar2023,Zhang2023,Bright2023,Gill2023}. 

It is important to note that none of these events share a significant gamma-ray brightness, as discussed in detail in Section~\ref{sec:gammasim}. This suggests that the optical emission and the gamma ray emission have different origins. Moreover, this illustrates the need to make multi-frequency observations for a complete understanding of each GRB.

The information provided in Table~\ref{tab:uvotbright} suggests that GRB~210822A and the extremely bright optical LGRBs considered in this work, share an intrinsically high isotropic energy ($E_\mathrm{iso}$) and luminosity ($L_\mathrm{iso}$). This, combined with the similarity found in their optical photometry, leads us to assume i) more extreme central engine properties which are thought to originate from a binary-driven hypernova produced by a binary system composed of a carbon-oxygen star and a neutron star unlike the conventional GRBs \citep{Rueda2020} or, ii) a specific configuration for the observer of a structured jet \citep{Lan2023}.

That said, \citet{Lan2023} compared the host galaxies of these intense GRBs with more conventional LGRBs and examined the energy, and they revealed no significant differentiation between these event groups.

The most accepted scenarios discussed for progenitors of LGRBs are a rapidly accreting black hole \citep[BH,][]{Narayan1992, Woosley1993, Fryer1999} and a rapidly spinning, strongly magnetised neutron star (NS) \citep[a millisecond magnetar;][]{Duncan1992, Usov1992, Thompson1994}. In the first scenario, the collapse of massive stars typically leads to a BH plus a long-lived debris torus system \citep[e.g., see][]{Woosley1993}. The relativistic jet is powered by extraction mediated by accretion from a Kerr BH \citep{Blandford1977}. When the star collapses into a BH, the stellar core must retain enough angular momentum to spin fast enough to generate an accretion disk.  In the last scenario, a millisecond magnetar is generated with a large rotational energy to impede the gravitational collapse \cite[e.g., see][]{Metzger2011}. The relativistic outflow is powered by the spin-down luminosity of the magnetic dipole, and the maximum energy reservoir becomes $\approx 2\times 10^{52}\,{\rm erg}\,M_{\rm ns}^{\frac32}P^{-2}_{-3}$ with $M_{\rm ns}$ the mass of the NS and $P$ the corresponding spin period.  Multiwavelength afterglow observations during the early phases provide valuable insights into elucidating the characteristics of the progenitors, the mass-loss evolution, metallicity, and establishing limits on the density of the surrounding medium \citep{Ackermann2013, Fraija2019, Fraija2020}. For instance, metallicity plays a relevant role in wind mass loss, along with rotation, determining the final phase angular momentum and, therefore, the destiny of the massive star. A millisecond magnetar can hardly explain a burst with isotropic energy larger than $10^{53}\,{\rm erg}$ and disfavor bursts that exhibit an optical flare due to a large magnetisation that would suppress the reverse shock \citep[e.g., see][]{Zhang2005}. Otherwise, highly bright LGRBs favour massive star progenitors with low metallicity \citep{Perley2014, Zhang2018}. 

%play a crucial role in clarifying the physical processes and emitting places associated with the spectral and temporal features of bursts. 

\section{Conclusions}
\label{sec:summary}

We have presented optical photometry of the afterglow of the ultra-bright GRB~210822A, combining new data from the RATIR instrument from $T+0.02$ to $T+0.49$~hours after the {\itshape Swift} trigger and with photometry from other telescopes published in GCN/TAN alerts.

We have compared the optical light curve and the X-ray obtained from {\itshape Swift}/XRT and {\itshape Swift}/UVOT, respectively. The photometric data reveal an achromatic temporal evolution well modelled by the expected evolution of an afterglow. The spectral indices measured at $T+500$ and $T+43000$ suggest the spectral regime where $\nu_\mathrm{m}<\nu_\mathrm{opt}<\nu_{X}<\nu_c$ corresponds to a flux evolving as $F\propto\nu^{\frac{p-1}{2}}t^{\frac{3(1-p)}{4}}$ with $p=2.54 \pm 0.10$.

Furthermore, we identified a reverse shock component in the optical bands before $T+300$~s. We also see a change in slope between $T + 80000$~s and $T + 100000$~s that we interpret as the signature of a jet break, allowing us to estimate a jet opening angle of about $\theta_\mathrm{j}=3-5^{\circ}$.

We use machine-learning techniques to model the light curves and constrain the intrinsic parameters of the burst. The technique used in this work is able to learn directly from the data and correct the model to use for fitting. This approach mitigates the challenges associated with conventional spline or polynomial interpolation methods and avoids excessive overfitting treatments.

In a general context, we showed the similarity of GRB~210822A with other events that are extremely bright in the optical/UV such as GRB~080319B, GRB~160625A GRB~190114C, GRB~221009A and GRB~130427A, which allows us to place GRB210822A in a peculiar group of GRBs with remarkably bright afterglows. Nevertheless, this set of GRBs differs from the events that share features in gamma-rays, evidencing the different regions and emission processes in which each of these frequencies originate. 

Finally, we underscore the importance of early-to-late ground-based telescope observations following a trigger, coupled with the essential supplementation of optical photometry data from {\itshape Swift}/XRT and {\itshape Swift}/UVOT. This combined approach is indispensable to achieve a comprehensive understanding of the GRB phenomena. Using earlier observations, we were able to model the reverse shock component and determine the level of low magnetisation present in this GRB. On the other hand, later observations constrain the opening angle and, therefore, the geometry of the jet, providing us with clues about the type of progenitor that produced GRB~210822A.

\section*{Acknowledgements}

The authors would like to thank the anonymous reviewer for his/her valuable comments.
Some of the data used in this paper were acquired with the RATIR instrument, funded by the University of California and NASA Goddard Space Flight Center, and the 1.5-meter Harold L. Johnson telescope at the Observatorio Astronómico Nacional on the Sierra de San Pedro Mártir, operated and maintained by the Observatorio Astronómico Nacional and the Instituto de Astronomía of the Universidad Nacional Autónoma de México. Operations are partially funded by the Universidad Nacional Autónoma de México (DGAPA/PAPIIT IG100414, IT102715, AG100317, IN109418, IG100820, IN106521 and IN105921). FDC acknowledges the computing time granted by DGTIC
UNAM on the supercomputer Miztli (project LANCAD-UNAM-DGTIC-281). We acknowledge the contribution of Leonid Georgiev and Neil Gehrels to the development of RATIR.

The authors thank the staff of the Observatorio Astronómico Nacional.

We acknowledge the support from the DGAPA/PAPIIT grants IG100422 and IN113424.

This work made use of data supplied by the UK Swift Science Data Centre at the University of Leicester. 

CAV, KGC, and FV acknowledge support from CONAHCyT fellowship. RLB acknowledges support from the CONAHCyT postdoctoral fellowship.
%%%%%%%%%%%%%%%%%%%%%%%%%%%%%%%%%%%%%%%%%%%%%%%%%%

\section*{Research Data Policy}
%\section*{Data Availability}
The data underlying this article will be shared on reasonable request to the corresponding author. The {\itshape Swift}/BAT and {\itshape Swift}/XRT data presented in this work are public and can be found in the domain \url{https://www.swift.ac.uk/xrt$\_$products/index.php}.\\



\bibliographystyle{mnras}
\bibliography{references} % if your bibtex file is called example.bib

%\begin{thebibliography}{99}
%\end{thebibliography}

%%%%%%%%%%%%%%%%%%%%%%%%%%%%%%%%%%%%%%%%%%%%%%%%%%

%%%%%%%%%%%%%%%%% APPENDICES %%%%%%%%%%%%%%%%%%%%%
\newpage
\appendix

%\section{Optical Photometry}
%\label{sec:Photometry}

\section{Neural Network Implementation}
\label{sec:neuronal}

In this study, we presented a machine learning approach that harnesses the power of 30,000 synthetic GRBs models to automatically generate models from observational data. While the generation of these synthetic datasets may be time-intensive, the training of our machine learning model is remarkably efficient. Furthermore, the application of the model to new observational data is even faster. A noteworthy advantage of this methodology is its versatility, as the trained ML model can be applied to a wide range of observational data, provided that the GRB models in question fall within the parameter space covered by the 30,000 synthetic models. This approach offers a powerful and expedient tool for modeling and analyzing GRBs, with the potential to significantly accelerate research in this field.

The implemented neural network architecture consists of a deep feedforward neural network with distinct layers. It incorporates multiple hidden layers and follows a pattern of stacking fully connected (dense) layers with activation functions, including ReLU and linear. Dropout layers are incorporated at various points within the architecture, serving as a form of regularisation to mitigate the risk of overfitting.

The final layer produces the output of the network output, producing predictions based on the initial input. This architecture allows the network to learn and model complex relationships within the data, contributing to its predictive capabilities. The idea is that we train the code with synthetic light curves as inputs and the eight parameters as outputs, so we can use the trained model to find parameters based on observed light curves.

\begin{table}%[h]
\centering
\begin{tabular}{|c|c|}
\hline
\hline
\text{Parameter} & \text{Range} \\
\hline
\hline
$\theta_\mathrm{obs}$ & $0.001$ to $0.150$~rad \\
$E$ & $10^{53}$ to $10^{55}$~erg\\
$\theta_\mathrm{j}$ & $0.001$ to $0.100$~rad \\
$p$ & $2.01$ to $3.00$ \\
$\epsilon_\mathrm{e}$ & $10^{-7}$ to $10^{-1}$ \\
$\epsilon_\mathrm{B}$ & $10^{-7}$ to $10^{-1}$ \\
$\chi$ & $0.8$ to $1.0$ \\
$n_\mathrm{0}$ & $0.7$ to $1.0$~cm$^{-3}$\\
\hline
\end{tabular}
\caption{Interval of each parameter used in the synthetic data generation.}
\label{table:parameter_ranges}
\end{table}

\begin{figure}
\includegraphics[width=0.9\linewidth]{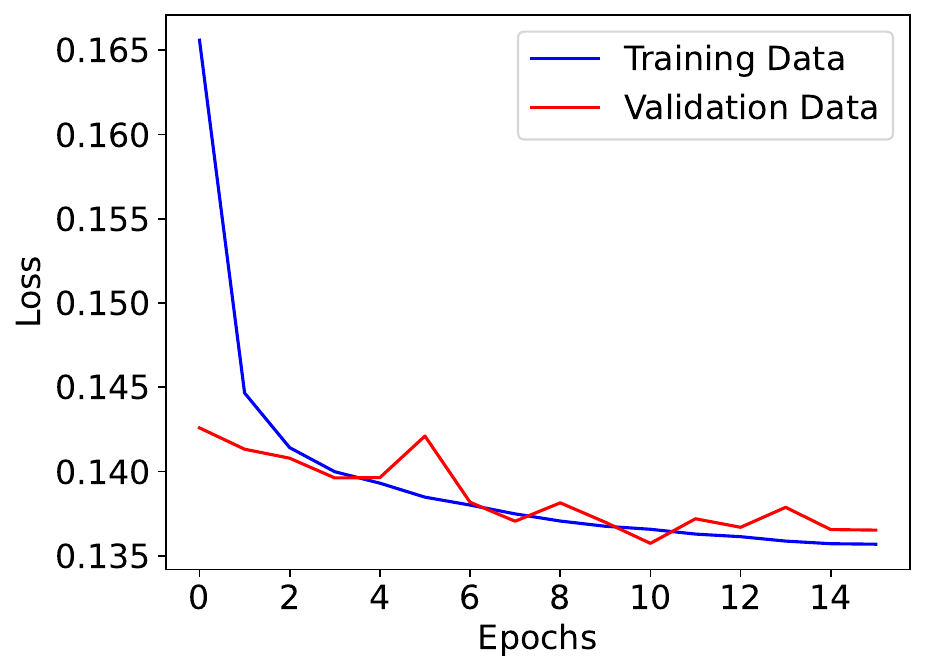}
 \caption{Temporal evolution of the loss function for training (blue) and validation data (red)}
 \label{fig:loss_epochs}
\end{figure}

% FABIO: la parte de ML es en un paper de felipe aun no publicado 
%\textbf{Analogously to what is described by \citet{Vargas2022}}, 
The ML model was trained over a span of 14 epochs (Figure \ref{fig:loss_epochs}). 
The number of training epochs is automatically determined through the implementation of an early stopping mechanism, the {\sc EarlyStopping} function \footnote{\url{https://www.tensorflow.org/api_docs/python/tf/keras/callbacks/EarlyStopping}}. This mechanism monitors the training process and stops training when convergence is detected in the loss function, facilitating efficient training by preventing overfitting and unnecessary iterations. The dataset was partitioned into three distinct sets: the training data, consisting of 25000 synthetic data points; the validation data, comprising 3000 samples; and the test data, with a total of 2000 samples. During each epoch, the ML model attempted to validate its learning by comparing its predictions against the validation dataset, which is not used in the training process. The primary objective of the training process is to minimise the loss function, the Mean Squared Error (MSE).

Once the model is adequately trained, it is evaluated against the test data set, which is used to estimate the errors in the parameters and to verify the absence of overfitting, which would restrict the neural network to reproducing only the training and validation data. Figure \ref{fig:loss_epochs} presents the temporal evolution of the loss function for both training and validation data.

% Don't change these lines
\bsp	% typesetting comment
\label{lastpage}
\end{document}

%% file: observations_RATIR.tex
\begin{table*}
	\centering
	\caption{Photometry of GRB 210822A.} %\textcolor{purple}{Faltan los errores de algunos datos de RATIR} 
	\label{tab:observations}
 \begin{tabular}{lcccl}
    \hline
    \hline
        \multicolumn{5}{c}{\bf {\itshape Data from this work (RATIR data)}} \\
    \hline
    \hline
			Time [s] & Exp [s]& Filter & Magnitude & Instrument\\
    \hline
    %\hline               
1199.2	&	80	&	\emph{ r } 	&	15.15	$\pm$	0.01	&	RATIR\\
1604.0	&	80	&	\emph{ r } 	&	15.53	$\pm$	0.01	&	RATIR\\
1709.0	&	80	&	\emph{ r } 	&	15.60	$\pm$	0.01	&	RATIR\\
\hline					
315.9	&	80	&	\emph{ i } 	&	13.17	$\pm$	0.00	&	RATIR\\
416.9	&	80	&	\emph{ i } 	&	13.78	$\pm$	0.00	&	RATIR\\
516.3	&	80	&	\emph{ i } 	&	14.15	$\pm$	0.00	&	RATIR\\
609.8	&	80	&	\emph{ i } 	&	14.20	$\pm$	0.00	&	RATIR\\
708.9	&	80	&	\emph{ i } 	&	14.47	$\pm$	0.00	&	RATIR\\
904.1	&	80	&	\emph{ i } 	&	15.00	$\pm$	0.00	&	RATIR\\
997.9	&	80	&	\emph{ i } 	&	14.93	$\pm$	0.01	&	RATIR\\
1098.6	&	80	&	\emph{ i } 	&	15.03	$\pm$	0.01	&	RATIR\\
1398.6	&	80	&	\emph{ i } 	&	15.34	$\pm$	0.01	&	RATIR\\
1604.0	&	80	&	\emph{ i } 	&	15.79	$\pm$	0.01	&	RATIR\\
1709.0	&	80	&	\emph{ i } 	&	15.86	$\pm$	0.01	&	RATIR\\
1803.8	&	80	&	\emph{ i } 	&	15.72	$\pm$	0.01	&	RATIR\\
    \hline
& & & &
%& & & &
 \end{tabular}

 \begin{tabular}{lcccll}
   \hline
   \hline
        \multicolumn{6}{c}{\bf {\itshape GCN data from optical ground-based facilities}} \\
    \hline
	\hline
			Time [s] & Exp [s]& Filter & Magnitude & Instrument & Reference \\
    \hline
    %\hline
114.0 & 20.0 & \emph{ w } & 11.40 $\pm$ $\cdot\cdot\cdot$ & KAIT & \cite{30679} \\
2690.0 & 20.0 & \emph{ w } & 16.10 $\pm$ $\cdot\cdot\cdot$ & KAIT & \cite{30679} \\
20424.9 & 1500.0 &  \emph{ r } & 18.53 $\pm$ 0.06 & NANSHAN & \cite{30684}\\
23760.0 & 10000.0 &  \emph{ V } & 18.88 $\pm$ 0.08 & NANSHAN & \cite{30684}\\
31262.9 & 3920 &  \emph{ R } & 18.63 $\pm$ 0.03 & TSHAO & \cite{30712}\\
37400.8 & 420.0  &  \emph{ i } & 18.56 $\pm$ 0.14 & CAHA & \cite{30723}\\
50004.0 & 360.0 &  \emph{ r } & 19.71 $\pm$ 0.02 & NOT & \cite{30692}\\
58665.6 & 2760.0 &  \emph{ R } & 20.10 $\pm$ 0.30 & Abbey Ridge & \cite{30701}\\
63822.0 & 600.0 &  \emph{ i' } & 19.92 $\pm$ 0.05 & GROND & \cite{30703}\\
63822.0 & 600.0 &  \emph{ g' } & 20.37 $\pm$ 0.09 & GROND & \cite{30703}\\
63822.0 & 600 &  \emph{ r' } & 20.10 $\pm$ 0.05 & GROND & \cite{30703}\\
201347.4 & 3180.0 &  \emph{ R } & 21.60 $\pm$ 0.26 & TSHAO & \cite{30714}\\
%&&&&& \\
\hline
 \end{tabular}
\end{table*}

%% file: similar.tex
% GRB & T90 & z & Refs
GRB 050128\tablefootnote{Information about duration is from the BAT catalogue \url{https://swift.gsfc.nasa.gov/results/batgrbcat/summary_cflux/summary_general_info/summary_general.txt}}  & $28.0\pm 9.1$ & 1.67 &  1 & $2.04$\\ % El t90 lo saqué de la página de swift, redshift no lo encuentro
GRB 070508  & $20.9\pm 0.7$ & 0.82   & 2, 3 & $2.34$   \\
GRB 080605  & $18.1\pm 0.9$ & 1.64   & 4   & $1.58$    \\ %no redshift error available
GRB 121128A & $23.4\pm 1.7$ & 2.20   & 5, 6 & $1.16$   \\ % no redshift error
GRB 140428A & $17.4\pm 5.9$ & 4.70   & 7, 8 & $2.19$  \\
\hline% no redshift error

%% file: uvot_r.tex
GRB 061007  &  $75.7 \pm 2.5$   & $\sim1.0\times10^{54}$  & $\sim1.8\times10^{53}$ & 1.26 & 0.8 - 4.7 & 1--6\\
GRB 080319B &  $124.9 \pm 3.1$  & $\sim1.3\times10^{54}$ & $\sim9.7\times10^{52}$ & 0.94 & 0.2 - 4.6 & 7--11\\
GRB 081203A &  $223.0 \pm 89.9$ & $\dots $ & $\dots$ & 2.05 &   \dots   & 12\\
GRB 130427A &  $244.3 \pm 4.7$  & $\sim8.5\times10^{53}$ & $\sim2.7\times10^{53}$ & 0.34 & 2.5 - 7.0 &  13--18\\
GRB 160625B$^{\rm *}$ & $460.0 \pm$ $\dots$ & $\sim5.0\times10^{54}$ & $\sim1.6\times10^{54}$ & 1.41 & 2.0 - 12 &  19--25\\
GRB 190114C & $361.5 \pm 11.7$  & $(2.4\pm0.1)\times10^{53}$ & $(1.7\pm0.1)\times10^{53}$ & 0.42 & 7.0 - 32  & 26--29\\
GRB 221009A & $1068.4 \pm 13.3$ & $(1.2\pm0.1)\times10^{55}$ & $(3.4\pm0.5)\times10^{54}$ & 0.15 & 0.7 - 2.0 & 30--35\\
GRB 210822A & $185.8 \pm 46.6$  & $(9.5\pm0.8)\times10^{53}$ & $(7.0\pm0.6)\times10^{53}$ & 1.74 & 3.0 - 5.0 & 36--39\\
\hline